\documentclass[12pt]{article}
\usepackage{amstex,epsfig}
\def\g5{\gamma^5}

\def\d4k{{d^4k\over (2\pi)^4}}

\newcommand{\beq}{\begin{eqnarray}}
\newcommand{\eeq}{\end{eqnarray}}
\newcommand{\beqno}{\begin{eqnarray*}}
\newcommand{\eeqno}{\end{eqnarray*}}
\def\lsim{\mathrel{\rlap{\lower4pt\hbox{\hskip1pt$\sim$}}
    \raise1pt\hbox{$<$}}}         
\def\gsim{\mathrel{\rlap{\lower4pt\hbox{\hskip1pt$\sim$}}
    \raise1pt\hbox{$>$}}}         

\begin{document}
%
\title{Meson Cloud, QCD Nucleon Correlator and Chiral Perturbation Theory}
 
\author{ Leonard S. Kisslinger\\
         Department of Physics,\\
       Carnegie Mellon University, Pittsburgh, PA 15213}
\maketitle
\begin{abstract}
We introduce a model for nucleons with correlators that include the 
propagation of pseudo Goldstone bosons as well as the standard correlators 
for nucleons used in QCD sum rules. From the comparison with experimental 
nucleon magnetic dipole moments we estimate the probability of a pion cloud 
in a nucleon. The masses obtained from the sum rules are almost independent 
of the cloud contributions. We show that the model is consistent with chiral
perturbation theory, and solves a long-standing problem on obtaining the
leading non-analytic correction to the nucleon mass in QCD sum rules.
\end{abstract}
\vspace{2 cm}

\indent
PACS Indices: 12.38.Lg,14.20Dh,13.40.Fn,11.50.Li

%
%
\newpage
\section{Introduction}
\hspace{.5cm}

   The concept of a meson cloud is very old in nuclear physics: the 
existence of the pion exchange interaction leads
to a pion cloud\cite{de} which could affect the properties of nucleons.
Although a main current goal of hadronic physics is to understand the
properties of hadrons in terms of QCD, in recent years the study of deep 
inelastic scattering (DIS) has led to a great deal of interest in the 
nucleon's meson cloud. The NMC/CERN experiments\cite{nmc} gave evidence
for the violation of the Gottfried sum rule and the recent 
Fermilab/E866\cite{fmlab1,fmlab2} 
Drell-Yan measurements show that for small Bjorken x the ratio
$\bar{d}(x)/\bar{u}(x)$, down to up sea quark distributions, is considerably 
larger than 1.0, while both perturbative\cite{rs} and nonperturbative 
QCD calculations\cite{jk}
show that $\bar{d}(x)/\bar{u}(x)$ for small x differs from unity only by
about 1 $\%$ isospin violations. It is likely that theoretical treatments
with both QCD and Goldstone boson (GB) fields are most efficient in treating
these sea quark problems as well as problems with the spin and strangeness
content of the nucleon. The challenge is to do this consistently.

   There have been many theoretical calculations of DIS using the concept of
a meson cloud based on the one-pion interaction of nucleons\cite{s} 
(see Ref.\cite{st} for a review). It is obvious that if one only considers
the $\pi^+$-neutron system as an additional component of the proton, that
there are more $\bar{d}$ than $\bar{u}$ sea quarks, but the $\pi^-\Delta$ 
components tend to cancel this effect. A related model is the chiral quark 
model which has been used for flavor and spin properties of 
nucleons\cite{cl}. As discussed in Ref.\cite{cl}, 
although these two models both introduce Goldstone
bosons, in the chiral quark model these are fields internal to the confinement
region, while the meson cloud model describes long-distance effects outside
of the region of confined quarks. In lattice gauge calculations\cite{liu}
of the flavor-singlet axial coupling constant for the nucleon's
quark spin content gluonic effects were included by disconnected
quark loops, which are interpreted as representing Goldstone bosons. In this
method valence quark and GB effects are distinguished. 

   It was observed that the QCD sum rules for the nucleon mass are
inconsistent with chiral perturbation theory\cite{gc}. First, the quark
condensates terms on the QCD side of the sum rules contain quark condensate 
terms which have been shown\cite{nsvz} to introduce $m_\pi^2 ln(-m_\pi^2)$
terms, while the chiral perturpation theory expansion for the nucleon 
mass\cite{gz} does not have such terms. Second, the leading non-analytic
correction (LNAC) to the nucleon mass is a m$_\pi^3$/f$_\pi^2$ term, which
is not found in QCD sum rules\cite{gc}. By an analysis of the 
phenomenological correlator, given by a term with a pole at the nucleon mass 
and the continuum, it was shown\cite{lccg} that the chiral logs cancel;
however, the LNAC term is not present so
the QCD sum rules for the nucleon mass are not consistent with chiral
perturbation theory. Virtual pions have
also been included in QCD sum rule calculations of nucleon masses in nuclear
matter\cite{dl,bi}. A chiral quark model was used to study pion-nucleon
weak and strong coupling within the QCD sum rule method\cite{ph}

    The model introduced in the present work uses the method of QCD Sum 
Rules\cite{svz} in which hadronic systems are represented by local complex 
field operators, usually called currents, so that two-point functions can 
be used for defining correlators from which one can determine hadronic
masses.  Nonperturbative effects are treated via operator product expansions 
(O.P.E.) of the quark and gluon propagators using vacuum condensates, whose 
values are determined by fits to  experiment as well as lattice gauge 
calculations. In this method  short-distance effects are treated 
perturbatively and the condensates account for the medium-range and long-range
QCD effects within the region of confinement.

  The model introduced in the present work treats the nucleon correlator
with a nucleon current that has a quark/gluon component and a component 
also involving a Goldstone boson field.
The GB fields are included in the composite-field  
nucleon current, with no coupling to either hadrons or quarks and thereby 
represent the long-distance meson cloud effects. Thus the model differs from 
both the usual meson cloud and the chiral quark models. All internal meson
exchange effects are assumed to be included in the condensates. For 
consistency, the coupling between the two currents is also not included, so
the correlator has a standard QCD sum rule part and a QCD/GB part, with a
GB propagator. There is a clear relationship between the present model of
nucleons with a meson cloud and the usual QCD sum rule treatment. In the
usual treatment states with a nucleon and a pion are in the continuum. In the
present model the current explicitly contains pionic fields, so that the
resulting continuum is modified. By doing this we are able to treat 
physical effects where explicit pion degrees of freedom are important, as
in sea-quark distributions.
   
   The goals of the present work are to estimate the probability of the meson
cloud component by studying the properties of the nucleon, and to study
the problem of consistency with chiral perturbation theory. This is
similar to the problem of finding the correlator for hybrid hadrons\cite{kl},
where the hybrid components of the nucleon and lowest hybrid baryon were
estimated from the hadronic masses. In Sec. 2 we introduce the correlator
using a standard form\cite{i} for the nucleon with no GB and derive the
sum rules for the nucleon masses including a correlator with a GB.
We find very small effects from the
meson cloud correlator terms. In Sec. 3 we derive the sum rules for the
neutron and proton magnetic dipole moments using
a two-point correlator in an external electromagnetic field\cite{is}. 
From the sum rules found with this method 
compared to QCD sum rule calculations of Ref.\cite{is} without the cloud,
we conclude that the GB component has roughly the same probability as
the cloudless component. It should be noted that
there is recent work on including instanton effects on the quark propagator 
that also modify the sum rules for the nucleon moments\cite{aw}.
In Sec. 4 we show that the model is consistent with chiral perturbation theory.
The chiral logs cancel and the LNAC term appears as in hadronic models.
The known value of the LNAC term from chiral perturbation theory can help
determine the parameters of the model.

\newpage
\section{Correlator with Goldstone Boson Fields: Nucleon Masses }
\hspace{.5cm}

  In this subsection we present our model of the pion cloud within the QCD
sum rule method, which means that we introduce a correlator with explicit
Goldstone boson fields. We apply this model correlator, a two-point function, 
to the estimate of the masses of nucleons.  
The composite field operator for the proton has the form
\beq
\eta(x) & = & c_1\eta^{p,o} + c_2\eta^{p,\pi} +  c_3\eta^{p,K},
\label{eq-eta}
\eeq
with constants c$_j$ giving the meson cloud amplitudes. In Eq.(\ref{eq-eta})
$\eta^{p,o}$ is a standard\cite{i} proton field operator used in sum rules:
\beq
\eta^{p,o}(x) & = &\epsilon^{abc}[u^a(x)^T C\gamma^\mu u^b(x)] \gamma^5
\gamma_\mu d^c(x),
\nonumber\\ 
 <0|\eta^{p,o}(x)|proton> & = & \lambda_p v(x),
\label{eq-etao}
\eeq
where C is the charge conjugation operator, the u(x) and d(x) are quark
fields labelled by color, $\lambda_p$ is a structure parameter and $v$(x)
is a Dirac spinor. The neutron operator, $\eta^{n,o}$, is obtained
from $\eta^{p,o}$ by interchange of u and d quarks.
The composite field operator $\eta^{p,\pi}$ used in our present model is
\beq
 \eta^{p,\pi}(x) & = & \frac{1}{\lambda_\pi^2}\partial_\alpha \phi_\pi(x)\cdot
\tau \gamma^\alpha \gamma^5 \eta^{N,o} \nonumber \\
                 & = &  \frac{1}{\sqrt{3}\lambda_\pi^2} 
[\sqrt{2} \partial_\alpha 
\phi_\pi^+(x) \gamma^\alpha \gamma^5 \eta^{n,o} + \partial_\alpha 
\phi_\pi^o(x) \gamma^\alpha \gamma^5 \eta^{p,o}], 
 \nonumber \\
  <0|\eta(x)^{p,\pi}|proton> & = &  \lambda'_p v(x),
\label{eq-etapi}
\eeq
where $ \phi_\pi(x)$ is the pion field and $\lambda_\pi$ is a D=1 scale 
factor.  From our experience with hybrid
baryons\cite{kl} and from the p-wave coupling of the pion one expects that
$\lambda^{'2}_p \ll \lambda_p^2$. This simplifies the phenomenological
form for the correlator and enables us to obtain sufficient sum rules
to estimate the pion cloud part of the current. 
In the present work we do not consider the kaon cloud term which 
would enter through $\eta^{p,K}$, defined similarly to Eq.(\ref{eq-etapi}) 
but with K fields and strange baryon composite operators. 
For estimates of strangeness in nucleons, which is a natural application of 
our model, this term must be included.

Our resulting model correlator for the proton with a pion cloud is
\beq
\Pi^p(p) & = 
  & i\int d^4 e^{ix\cdot p}<0|T[\eta(x)\bar\eta(0)]|0> \nonumber\\
  & = & c_1^2 i\int d^4 e^{ix\cdot p}<0|T[\eta^{p,o}(x)\bar\eta^{p,o}(0)]|0>
  + (1-c_1^2)i\int d^4 e^{ix\cdot p}<0|T[\eta^{p,\pi}(x)
 \bar\eta^{p,\pi}(0)]|0> 
 \nonumber\\
  &=& c_1^2 \Pi^{(p,o)}(p) + (1-c_1^2)\Pi^{(p,\pi)}(p).
\label{eq-cor}
\eeq 
An important aspect of the model is that the coupling of the $\eta^{p,\pi}$
and $\eta^{p,0}$ currents are not included. Only the long-range effects
of the pion are included by the pion cloud contribution, and no pion-quark
interactions are included in order to avoid the double counting of 
interactions given by the condensates. In a chiral quark model the 
pion-quark coupling would give both the coupling of the two currents and
internal pionic forces. The present model assumes such internal interactions
are given by QCD. Only external fields interact with the pion, giving the
physics of the meson cloud. The neutron correlator can be obtained
from Eq.(\ref{eq-cor}) with obvious isospin changes.
The correlator $\Pi^{(p,0)}(p)$ for the propagation of a proton without
the meson cloud is given in Ref.\cite{is} and many other references.

Next we discuss the correlator with the pion cloud, $\Pi^{(p,\pi)}(p)$.
From Eqs.(\ref{eq-etapi},\ref{eq-cor}) one finds for  
 $\Pi^{(p,\pi)}(x)$ 
\beq
  \Pi^{(p,\pi)}(x) & = &  \frac{2}{3\lambda_\pi^4}\epsilon^{abc}
 \epsilon^{b'a'c'} G_{\alpha\beta}^{\pi}(x)
 \gamma_\alpha\gamma_\mu  S^{cc'}_d(x) \gamma_\nu \gamma_\beta  
 \nonumber \\ 
  & & Tr[S^{aa'}_u(x) \gamma^\mu C (S^{bb'}_u(x))^T C\gamma^\nu]
 \nonumber \\
  & & + 2 [\pi^0 \rightarrow \pi^+\  and\  u \leftrightarrow d] + 
 \ [4-quark\ terms],
\label{eq-corpi}
\eeq
where in the limit of zero pion mass 
\beq
 G_{\alpha\beta}^\pi(x) & = & \frac{1}{\pi^2 x^4}(\frac{4x_\alpha x_\beta}
{x^2} - \delta_{\alpha\beta}).
\label{G}
\eeq
In momentum space the pion cloud correlator is
\beq
  \Pi^{(p,\pi)}(p) & = & -2\int \frac{d^4k}{(2\pi)^4}\frac{(\hat{p}-\hat{k})
 \Pi^{p,o}(k)(\hat{p}-\hat{k})}{(p - k)^2} \nonumber \\
   & = & \hat{p} \Pi^{(p,\pi)}(p)_{odd} + \Pi^{(p,\pi)}(p)_{even},
\label{eq-corpip}
\eeq
with $\hat{p}$= $\gamma_\alpha p^\alpha$.
We find after the Borel transform from p$^2$ to the Borel mass variable,
M$_B^2$
\beq
   \Pi^{(p,\pi)}(M_B^2)_{even} & = & 0 \nonumber \\
  (2 \pi)^4 \Pi^{(p,\pi)}(M_B^2)_{odd}   & = & \frac{1}{\lambda_\pi^4 
 15\cdot 2^8} [12 M_B^{10}E_4 -5 b M_B^6 E_2 - 40 a^2 M_B^4 E_1 
 \nonumber \\
  & & +20 m_o^2 a^2 M_B^2 E_0],
\label{eq-corpim}
\eeq
where the parameters for the quark condensate, gluon condensate and mixed
condensate, respectively, are a = -$(2\pi)^2 <\bar{q}q>$, b = $ <\bar{q}
 g^2 G^2 q>$ and m$_o^2$ a = $(2\pi)^2 <\bar{q} g \sigma \cdot G q>$.
For these parameters we take a = .55 GeV$^3$, b = .47 GeV$^4$ and m$_o^2$ = 
0.8 GeV$^2$. The scale factor $\lambda_\pi^2$ has a value in the range of
M$_B^2/g_{\pi N}$ (see Sec. 4). The functions E$_n$ are 
1-exp(-w)$\sum_{k=0}^{n}$w$^k/k!$,
with w = s$_o$/M$_B^2$, s$_o$ being the threshold parameter.
In comparison with Eq.(\ref{eq-corpim}) for the meson cloud
terms in the correlator, the usual(p,0) odd correlator is
\beq
  (2 \pi)^4  \Pi^{(p,0)}(M_B^2)_{odd} & = & \frac{1}{2^5} (-4 M_B^6 E_2
 +b M_B^2 E_0 + 16 a^2/3 -4 m_o^2 a^2/(3 M_B^2).
\label{eq-corom}
\eeq
The dependence of the correlators on the anamoulous dimensions are not
shown in Eqs.(\ref{eq-corpim},\ref{eq-corom}), since the result is not 
essentially changed if this is neglected. In fact it is obvious from
Eqs.(\ref{eq-corpim},\ref{eq-corom}) that if $\lambda^{'2}_p << \lambda_p^2$,
as expected, then the sum rules for the masses are independent of the pion
cloud within the accuracy of the method.
 
\section{Correlator with Goldstone Boson: Nucleon Magnetic Dipole Moments}

\hspace{.5cm}

  For the calculation of the magnetic dipole moments of the proton and
neutron we use the external field formalism, which is one way to avoid the 
problem of carrying out an operator product expansion in the momentum transfer
variable. The coupling of a current $J^\Gamma$ to a nucleon is given by
a two-point correlator in the external  $J^\Gamma$ current\cite{is}
\beq
\Pi^\Gamma(p) & = 
  & i\int d^4 e^{ix\cdot p}<0|T[\eta(x)\bar\eta(0)]|0>_{J^\Gamma}.
\label{eq-gpi}
\eeq
For the proton in an  electromagnetic field, represented by the field tensor
F$_{\mu\nu}$ the correlator is written as
\beq
\Pi^{F,p} & =  & e (c_1^2 \Pi^{p,o}_{\mu\nu} + (1-c_1^2)\Pi^{p,\pi}_
{\mu\nu}) F^{\mu\nu},
\label{eq-piem}
\eeq
with the two terms representing the correlator in the electromagnetic
field without and with the meson cloud. Since at low momentum transfer
the coupling of the photon to the pion does not contribute, the meson
cloud correlator in the presence of the field in our model is given by
\beq
  \Pi^{(p,\pi)}_{\mu\nu}(p) & = & -2\int \frac{d^4k}{(2\pi)^4}
\frac{(\hat{p}-\hat{k}) \Pi^{p,o}_{\mu\nu}(k)(\hat{p}-\hat{k})}{(p-k)^2} 
\label{eq-corpipm}
\eeq
There are three covariants, and we write the correlator with and without
meson clouds as
\beq
\Pi_{\mu\nu}(k) & =  &  [\sigma_{\mu\nu},\hat{k}] \Pi^F_{odd}(k)
 + i(k_\mu \gamma_\nu -k_\nu \gamma_\mu) \Pi^F_{even}(k) 
 + \sigma_{\mu\nu} \Pi^F_\sigma (k).
\label{eq-cor2}
\eeq

   For the two-point effective field treatment at low momentum  momentum
transfer $\Pi^\Gamma(x)$ can be evaluated using the operator product 
expansion (O.P.E.),
since the variable x is at short distance from the origin for
large p$^\mu$. This is done by an
O.P.E. of the quark propagator in the presence of the the $J^\Gamma$ current
\beq
 S^\Gamma_q(x) & = & <0|T[q(x)\bar{q}(0)]|0>_{J^{\Gamma}}.
\label{eq-gprop}
\eeq 
There are three susceptibilities for the induced condensates: $\chi,
\kappa$ and $\xi$, which are defined in Ref\cite{is}. For example, the
quark condensate magnetic susceptibility, $\chi$, is defined as
\beq
 (2\pi)^2<\bar{q}\sigma_{\mu\nu}q>_F & = & -e_q \chi a F_{\mu\nu}.
\label{eq-chi}
\eeq

The QCD evaluation of the correlator in an electromagnetic field without 
a meson cloud gives\cite{is}
\beq
 (2\pi)^4 \Pi^{p,o,F}_{odd}(k) & =  & e_u k^2 ln(-k^2) - e_u \chi a^2/(3 k^2)
 +C_a/(3 k^4) \nonumber \\
 (2\pi)^4 \Pi^{p,o,F}_{even}(k) & =  & a[(e_u +e_d/2)/k^2 +
 (e_d \chi/3)( ln(-k^2) -b/(24 k^4)],
\label{eq-cor3}
\eeq     
with C$_a$ = $(a^2/2)[e_d + 2 e_d/3 -e_u(\kappa - 2\xi)] -e_u \chi a^2
m_o^2/8$.
From Eqs.(\ref{eq-corpim},\ref{eq-cor3}) we obtain the correlator for
the nucleon with a pion cloud in an external electromagnetic field, giving
\beq
 (2\pi)^4 \Pi^{p,\pi,F}_{odd}(p) & =  & \frac{ln(-p^2)}{3\cdot2^8 \pi^2}
 [- e_u p^6 + 40 e_u \chi a^2 p^2/3 +32 C_a] \nonumber \\
 (2\pi)^4 \Pi^{p,\pi,F}_{even}(k) & =  & - \frac{ln(-p^2)}{2^5\cdot3^3\cdot 5 
 \pi^2}\chi a[3^2\cdot2^4e_d p^4 - 5 e_d b].
\label{eq-cor4}
\eeq
For a rough estimate of the effect of the meson cloud, which is the
goal of the present paper, we follow the prescriptions of Ref.\cite{is}:
For the odd term in the correlator multiply the p term by e$_d$ and subtract
e$_u$ times the n term. Neglecting anamolous dimensions one finds
\beq
 \frac{-\bar{\lambda}^2_N}{4}e^{-M_N^2/M_B^2}(e_d \mu_p - e_u \mu_n)/M_B^2 +
 single \ poles & =& (e_d^2-e_u^2)[\frac{a^2}{6M_B^2} + \Delta_{cl}],
\label{eq-del}
\eeq
where $\bar{\lambda}^2_N$ = (2$\pi$)$^4 \lambda_N$ and the GB contribution is
\beq
\Delta_{cl} & = & \frac{1-c_1^2}{c1^2\lambda_\pi^4}\frac{1}{2^7 3^2 \pi^2} 
[8 a^2 M_B^2 E_0-6 M_B^8 E_3-\frac{40}{3} \chi a^2 E_1+
\frac{32}{3} a^2 E_0+\frac{1}{8}\chi a^2 m_o^2].
\label{eq-delmu}
\eeq
Noting that with a value of $\chi$ a = -4 GeV, which we obtain with our
three-point treatment with nonlocal condensates\cite{k}, a value of c$_1^2$
about 0.5 reduces the neutron and proton magnetic dipole moments by about 
ten percent, improving agreement with experiment.  From this we conclude that 
the component of the correlator with a pion cloud is roughly equal to that
without. As we shall see in the next section, the known LNAC for the nucleon
mass will help determine the scaling parameter, $\lambda_\pi$.

\section{Consistency of Model With Chiral Perturbation Theory}
 
\hspace{.5cm}

In Ref. \cite{nsvz} it was shown that from the dependence of the quark
condensates on the current quark masses and the Gell-Mann-Oakes-Renner
relation\cite {gor} that the quark condensate has a  $m_\pi^2 ln(-m_\pi^2)$
dependence on the pion mass. From this one can see from Eq. (\ref{eq-corom}
that the standard QCD sum rule contains chiral logs that are not consistent 
with chiral perturbation
theory\cite{gc}. Moreover, the LNAC term, which is proportional to
m$_\pi^3$/f$_\pi^2$ \cite{gz} does not occur in the QCD sum rule form for
the nucleon mass. in this section we show that both of these problems are
solved in our model.

\subsection{Chiral logarithms} 

\hspace{.5cm}

In this subsection we show that by using the 
results of
Ref. \cite{lccg} our present model does not contain inconsistent chiral
logarithms. We study the diagrams giving the LNAC in the next subsection.
For simplicity let us take $c_1^2$ = 0.5. Using the observation from
hybrid baryons\cite{kl} that  $\lambda^{'2}_p << \lambda_p^2$, as discussed
in Sec. 2, the phenonenological side of the correlator, given by a 
dispersion relation has the form of a pole term and a continuum term,
\beq
\label{cp1}
  2 \Pi^{p(Phen)}(p) & = & -\lambda_p^2 \frac{\hat{p} + M_p}{p^2 - M_p^2}
 +  \Pi^{p(cont)}(p),
\eeq
which is the same as the standard proton correlator except for a factor of 
two. As we saw in Sec. 2, the microscopic QCD side of the correlator has
the form
\beq
\label{cp2}
  2 \Pi^{p(qcd)}(p) & = &  \hat{p}( \Pi^{(p,0)QCD}(p)_{odd} +
 \Pi^{(p,\pi)QCD}(p)_{odd}) + \Pi^{(p,0)}(p)_{even}.
\eeq
Using the standard proton current\cite{i} we recall that
\beq
\label{cp3}
  \Pi^{(p,0)QCD}(p)_{even} & = &  <\bar{q}q> (p^2 ln(-p^2) - \frac{b}{18 p^2})
   + \cdots \\ \nonumber
  \Pi^{(p,0)QCD}(p)_{odd} & = &  -ln(-p^2) (\frac{p^4}{4} + \frac{b}{8})
   + \cdots,
\eeq
where by the ellipsis we mean four-quark condensates + higher dimensional 
terms. Making use of the $m_\pi$ expansion of the quark condensate from 
Ref. \cite{nsvz}:
\beq
\label{cp4}
    <\bar{q}q> & = &  <\bar{q}q>_o [1-c m_\pi^2ln(-m_\pi^2)]
\eeq
where $  c = \frac{3}{32 \pi^2 f_\pi^2}$ and by the symbol $Q_o$ 
we mean the quantity Q in the limit $m_\pi^2 \rightarrow 0$, one finds 
\beq
\label{cp5}
         \Pi^{(p,0)QCD}(p)_{even} & = & [1-c m_\pi^2ln(-m_\pi^2)]
 \Pi_o^{(p,0)QCD}(p)_{even} + O(m_\pi^2) 
 \\ \nonumber
          \Pi^{(p,0)QCD}(p)_{odd} & = & [1+O(m_\pi^2)] 
  \Pi_o^{(p,0)QCD}(p)_{odd},
\eeq
as was observed in Ref.\cite{lccg}. For the pion cloud part of the correlator
we observe that
\beq
\label{cp6}
    \Pi^{(p,\pi)}(x) & = &  G_{\alpha\beta}^{\pi}(x) \gamma^\alpha \Pi^{(p,0)}
 \gamma^\beta + m_\pi^2 \breve{G}_{\alpha\beta}^{\pi}(x) \gamma^\alpha \Pi^{(p,0)}
 \gamma^\beta,
\eeq
where $\breve{G}_{\alpha\beta}$ is obtained from ${G}_{\alpha\beta}$ of 
Eq.(\ref{eq-corpi}) by an expansion of the pion propagator in powers of
$m_\pi^2$, with the first term given by Eq(\ref{G}). From Eq.(\ref{cp6} we
observe that
\beq
\label{cp7}
  \Pi^{(p,\pi)}(x) & = &  \Pi^{(p,\pi)}_o(x) + O(m_\pi^2),
\eeq
and does not contain any $m_\pi^2 ln(-m_\pi^2)$ chiral log terms.

Therefore we conclude that in our model of the nucleon with a meson cloud
\beq
\label{cp8}
     \Pi^{p(Phen)} & = & [1-c m_\pi^2 ln(-m_\pi^2)]  \Pi^{p(Phen)}_o + \cdots
\eeq
and
\beq
\label{cp9}
     \Pi^{p(QCD)} & = & [1-c m_\pi^2 ln(-m_\pi^2)]  \Pi^{p(QCD)}_o + \cdots;
\eeq
and our model the nucleon with pion cloud does not contain spurious
chiral logarithms.

\subsection{Leading Non-Analytic Correction}

\hspace{.5cm}

The leading non-analytic correction (LNAC) from chiral symmetry breaking
can be found in chiral perturbation theory using the method described in
Refs.\cite{lp,gz}, with the Feynman-like diagrams of chiral perturbation
theory given in Ref.\cite{ch}. The main idea is that if one has a broken
symmetry, so that the Hamiltonian has the form
\beq
\label{cp10}
                H & = & H_0\ \ +\ \ m H_I,
\eeq
with m being the small parameter of the symmetry breaking, the mass of a
hadron can be found from the relation
\beq
\label{cp11}
           \frac{\partial^2 M^2}{\partial m^2} & = & Lim_{q^0 \rightarrow 0^+,
q^2 \rightarrow 0,{\bf p} \rightarrow 0}i\int d^4x e^{iqx}<p|\theta(x^o)
[\mathcal{H}_I(x),\mathcal{H}_I(0)]|p>,
\eeq
including only connected diagrams, with H$_I$ = $\int d^3x \mathcal{H}_I$.
The chiral symmetry breaking in QCD is
\beq
\label{cp12}
             m H_I & = & \int d^3x m_u \bar{u}u + m_d \bar{d}d \nonumber \\
                   & = & m \int d^3x \mathcal{H}_m.
\eeq

   In Ref \cite{gz} it is shown that the LNAC for the nucleon mass is
given by the process illustrated in Fig. 1.
\begin{figure}[htpb]
\begin{center}
\epsfig{file=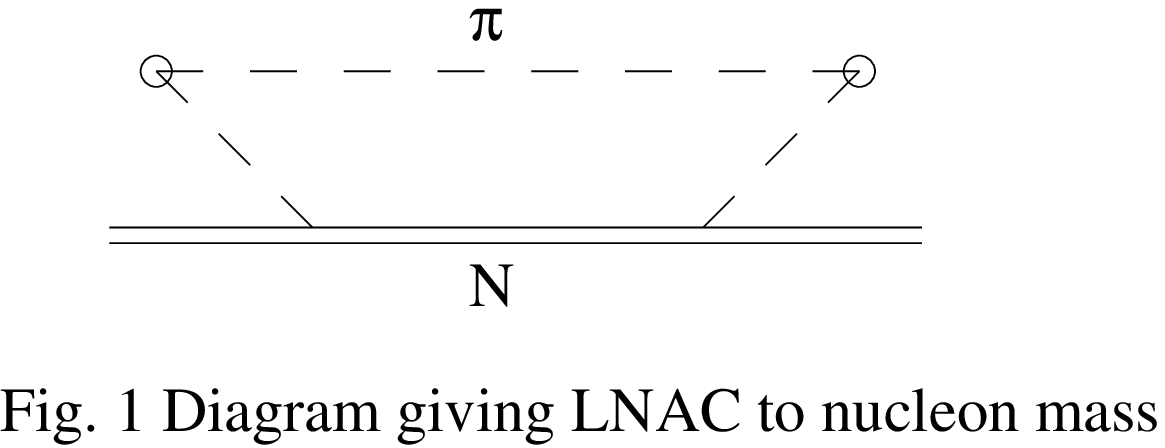,width=10cm}
{\label{Fig.1}}
\end{center}
\end{figure}
The corresponding correction to the nucleon mass, $\Delta M_N = M_N - M_N^o$
was shown to be
\beq
\label{cp13}
 \Delta M_N & = & -\frac{3 g_A^2 m_\pi^3}{32 \pi f_\pi^2} + \cdots \ \ = \ \
-15 MeV + \cdots,
\eeq
which makes use of the result
\beq
\label{cp14}
   <\pi^a|\mathcal{H}_m|\pi^b> & = & \delta_{ab} \frac{m_\pi^2}{m} + \cdots. 
\eeq      

The LNAC in the present model of QCD sum rules, after the chiral logs have
been eliminated, are given by processes involving the pion cloud correlator.
The lowest dimensional term is illustrated by the the diagram shown in
Fig. 2
\begin{figure}[htpb]
\begin{center}
\epsfig{file=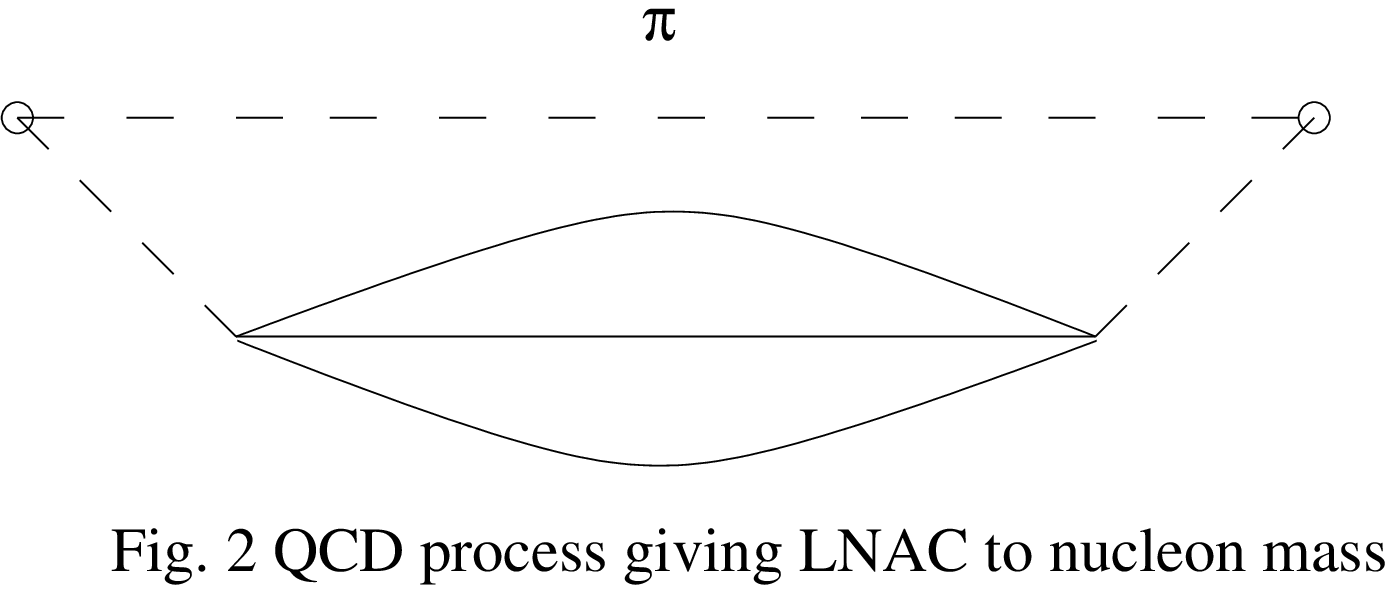,width=10cm}
{\label{Fig.2}}
\end{center}
\end{figure}
In the evaluation of the QCD sum rule diagram the $\pi$-N T-matrix evident in
Fig. 1 is replaced by the traces over the quark propagators. The LNAC to the
nucleon mass is found by equating the phenomenological expression for the
Borel transformed correlator
\beq
\label{cp15}
         \Pi^{p(Phen)}(p) & = & \beta^2 exp[-\frac{M_p^2}{M_B^2}] + continuum,
\eeq
with $\beta^2 = (2 \pi)^4 \lambda_p^2/4$ (see Eq.(\ref{eq-etao}), 
to the QCD expression with chiral symmetry breaking.  Making use of
Eq(\ref{cp14}), the Goldberger-Trieman relation 
(g$_{\pi N}$=g$_A$ M$_N$/f$_\pi$), Eq.(\ref{eq-corpim}) and taking the nucleon
and Borel masses M$_p$, M$_B$ = 1 Gev,
we find for the lowest-dimensional LNAC term
\beq
\label{cp16}
   \Delta M_p & = & -\frac{e m_\pi^3 g_A^2}{160 \beta^2 f_\pi^2} \nonumber \\ 
              & \simeq & -20 MeV.
\eeq
For the result of a LNAC of about 20 MeV, which is consistent with the chiral
perturbations theory result, we have used a value of $\beta^2$ from 
Ref. \cite{is}, where it is pointed out that there is about a 50\% 
uncertainty in the value, and have used  $\lambda_\pi^2$ = M$_p^2$/g$_{\pi N}$
for the scale factor of the pion cloud model.  Note that the gluon condensate 
process also contributes to the LNAC, with an opposite sign, cancelling  
about 10-20\% of the term shown in Eq. (\ref{cp16}). 
   
Therefore, we have shown that the longstanding puzzle of the absence of the
LNAC m$_\pi^3$ term in the QCD sum rule treatment of the nucleon mass can
be explained in our model, with a correlator composed about equally of
a conventional quark correlator and a meson cloud correlator. The known
value of the LNAC for the nucleon mass can help determine the parameters
of the model.

\section{Conclusions}
We have introduced a model for a nucleon in which Goldstone bosons are
included in a QCD treatment. By only including the boson propagator and
not interior quark-boson interactions we hope to account for long-distance
effects which are difficult to calculate in a pure QCD treatment, and yet
avoid introducing processes which are aleady treated by the nonperturbative
QCD methods of QCD sum rules. With this model the nucleon mass sum rules
are essentially not changed by the meson cloud terms, but from the
treatment of magnetic dipole moments we conclude that within our model
the correlators with and without the pion cloud are about equal. We have
also shown that this model is consistent with chiral perturbation theory,
and that the leading nonanalytic term found in chiral perturbation theory,
but missing in previous treatments of the nucleon mass with QCD sum rules,
is present in our model. The magnetic dipole moments and the LNAC for the
nucleon mass can help determine the parameters of the model.
Our future research with this model will include 
studies such as the spin and strangeness content of the proton, as well as
the sea-quark distributions in the nucleon.

  The author would like to thank Montaga Aw, Andrew Harey, Ling-fong Li
and Jen-Chieh Peng for helpful discussions.  This work was supported in 
part by NSF Grant PHY-9722143.

\hspace{.5cm}
 


\begin{thebibliography}{99}
\bibitem{de}S. DeBenedetti, ``Nuclear Interactions'' (John Wiley and Sons,
New York)(1964).
\bibitem{nmc}P. Amaudruz et. al., Phys. Rev. Lett, {\bf 66}, 2712 (1991);
 M. Arneodo et. al., {\bf D 50}, R1 (1994).
\bibitem{fmlab1}E. A. Hawker et. al., Phys. Rev. Lett. {\bf 80}, 3715 (1998).
\bibitem{fmlab2}J. C. Peng et. al., Phys. Rev. {\bf D 58}, 92004 (1998)
\bibitem{rs}D. A. Ross and C. T. Sachrajda, Nucl. Phys. {\bf B149}, 497
(1979).
\bibitem{jk} H. Jung and L. S. Kisslinger, Nucl. Phys. {\bf A586}, 682
(1995).
\bibitem{s}J. D. Sullivan, Phys. Rev. {\bf D 5}, 1732 (1972).
\bibitem{st} J. Speth and A. W. Thomas, Adv. Nucl. Phys. {\bf 24}, 83 (1998).
\bibitem{cl} T. P. Cheng and L.-F. Li, Phys. Rev. Lett. {\bf 74}, 2872 (1995);
 "Non-Perturbative QCD Spin Studies", hep-ph/9811279.
\bibitem{liu}K. F. Liu, Phys. Lett. {\bf B 281}, 141 (1992); S. J. Dong,
 J.-F. Laga\"{e} and K. F. Liu, Phys. Rev. Lett. {\bf 75}, 2096 (1995).
\bibitem{gc}D.K. Griegel and T.D. Cohen, Phys. Lett. {\bf B333}, 27 (1994).
\bibitem{nsvz}V.A. Novikov, M.A. Shifman, A.I. Vainstein, and V.I. Zakharov,
 Nucl. Phys. {\bf B191}, 301 (1991).
\bibitem{gz} J. Gasser and A. Zepeda, Nucl. Phys. {\bf B174}, 445 (1980).
\bibitem{lccg} S.H. Lee, S. Choe, T.D. Cohen and D.K. Griegel, Phys. Lett.
 {\bf B348}, 263 (1995); K. Maltman and M.C. Birse, Phys. Rev. {\bf C56}, 1588
 (1997).
\bibitem{dl}E. G. Drukarev and E. M. Levin, JETP Lett. {\bf 48}. 338 (1988);
Nucl. Phys. {\bf A511}, 679 (1990); Prog. Part. Nucl. Phys. {\bf 27}, 77 
(1991).
\bibitem{bi}M. C. Birse, Phys. Rev. {\bf C 53}, R2048 (1996); M. C. Birse and
B. Krippa, Phys. Lett. {\bf B 381}, 397 (1996).
\bibitem{ph}W-Y. P. Hwang, Z. Phys. {\bf C75}, 701 (1997).
\bibitem{svz} M. A. Shifman, A. I. Vainshtein, and V. I. Zakharov, Nucl.
Phys. {\bf B147} (1979) 385; 448.
\bibitem{kl}L. S. Kisslinger and Z. Li, Phys. Rev. Lett. {\bf 74}, 2168
 (1995).
\bibitem{i} B. L. Ioffe, Nucl. Phys. {\bf B188}, 317 (1981); V.M. Belyaev
 and  B.L. Ioffe Sov. Phys.JETP {\bf 56}, 493 (1982).
 \bibitem{is} B. L. Ioffe and A. V. Smilga, Nucl. Phys. {\bf B232}, 109
(1984).
\bibitem{aw}M. Aw, M. K. Banerjee and H. Forkel, Phys. Lett. {\bf B454},
 147 (1999).
\bibitem{k}L. S. Kisslinger,  Phys. Rev. {\bf C 59}, 3377 (1999).
\bibitem{gor}M. Gell-Mann, R.J. Oakes and B. Renner, Phys. Rev. {\bf 175},
 2195 (1968)
\bibitem{lp}P. Langacker and H. Pagels, Phys. Rev. {\bf D8}, 4595 (1973).
\bibitem{ch}J. M. Charap, Phys. Rev. {\bf D2}, 1554 (1970).

\end{thebibliography}
\end{document}